\input harvmac
%

\let\includefigures=\iftrue
\let\useblackboard=\iftrue
\newfam\black

\includefigures
\message{If you do not have epsf.tex (to include figures),}
\message{change the option at the top of the tex file.}
\input epsf
\def\figin{\epsfcheck\figin}\def\figins{\epsfcheck\figins}
\def\epsfcheck{\ifx\epsfbox\UnDeFiNeD
\message{(NO epsf.tex, FIGURES WILL BE IGNORED)}
\gdef\figin##1{\vskip2in}\gdef\figins##1{\hskip.5in}
\else\message{(FIGURES WILL BE INCLUDED)}%
\gdef\figin##1{##1}\gdef\figins##1{##1}\fi}
\def\DefWarn#1{}
\def\figinsert{\goodbreak\midinsert}
\def\ifig#1#2#3{\DefWarn#1\xdef#1{fig.~\the\figno}
\writedef{#1\leftbracket fig.\noexpand~\the\figno}%
\figinsert\figin{\centerline{#3}}\medskip\centerline{\vbox{
\baselineskip12pt\advance\hsize by -1truein
\noindent\footnotefont{\bf Fig.~\the\figno:} #2}}
\endinsert\global\advance\figno by1}
\else
\def\ifig#1#2#3{\xdef#1{fig.~\the\figno}
\writedef{#1\leftbracket fig.\noexpand~\the\figno}%
\global\advance\figno by1} \fi

\def\journal#1&#2(#3){\unskip, \sl #1\ \bf #2 \rm(19#3) }
\def\andjournal#1&#2(#3){\sl #1~\bf #2 \rm (19#3) }

\def\ie{{\it i.e.}}
\def\eg{{\it e.g.}}

\noblackbox
%


\def\unlockat{\catcode`\@=11}
\def\lockat{\catcode`\@=12}

\unlockat

\def\newsec#1{\global\advance\secno by1\message{(\the\secno. #1)}
\global\subsecno=0\global\subsubsecno=0\eqnres@t\noindent
{\bf\the\secno. #1}
\writetoca{{\secsym} {#1}}\par\nobreak\medskip\nobreak}
\global\newcount\subsecno \global\subsecno=0
\def\subsec#1{\global\advance\subsecno
by1\message{(\secsym\the\subsecno. #1)}
\ifnum\lastpenalty>9000\else\bigbreak\fi\global\subsubsecno=0
\noindent{\it\secsym\the\subsecno. #1}
\writetoca{\string\quad {\secsym\the\subsecno.} {#1}}
\par\nobreak\medskip\nobreak}
\global\newcount\subsubsecno \global\subsubsecno=0
\def\subsubsec#1{\global\advance\subsubsecno by1
\message{(\secsym\the\subsecno.\the\subsubsecno. #1)}
\ifnum\lastpenalty>9000\else\bigbreak\fi
\noindent\quad{\secsym\the\subsecno.\the\subsubsecno.}{#1}
\writetoca{\string\qquad{\secsym\the\subsecno.\the\subsubsecno.}{#1}}
\par\nobreak\medskip\nobreak}

\def\subsubseclab#1{\DefWarn#1\xdef
#1{\noexpand\hyperref{}{subsubsection}%
{\secsym\the\subsecno.\the\subsubsecno}%
{\secsym\the\subsecno.\the\subsubsecno}}%
\writedef{#1\leftbracket#1}\wrlabeL{#1=#1}}
\lockat

\def\ie{{\it i.e.}}
\def\eg{{\it e.g.}}


\font\manual=manfnt \def\dbend{\lower3.5pt\hbox{\manual\char127}}

\def\IZ{\relax\ifmmode\mathchoice
{\hbox{\cmss Z\kern-.4em Z}}{\hbox{\cmss Z\kern-.4em Z}}
{\lower.9pt\hbox{\cmsss Z\kern-.4em Z}}
{\lower1.2pt\hbox{\cmsss Z\kern-.4em Z}}\else{\cmss Z\kern-.4em
Z}\fi}
\def\half{{1\over 2}}


\def\IZ{\relax\ifmmode\mathchoice
{\hbox{\cmss Z\kern-.4em Z}}{\hbox{\cmss Z\kern-.4em Z}}
{\lower.9pt\hbox{\cmsss Z\kern-.4em Z}}
{\lower1.2pt\hbox{\cmsss Z\kern-.4em Z}}\else{\cmss Z\kern-.4em
Z}\fi}
\def\IB{\relax{\rm I\kern-.18em B}}
\def\IC{{\relax\hbox{$\inbar\kern-.3em{\rm C}$}}}
\def\ID{\relax{\rm I\kern-.18em D}}
\def\IE{\relax{\rm I\kern-.18em E}}
\def\IF{\relax{\rm I\kern-.18em F}}
\def\IG{\relax\hbox{$\inbar\kern-.3em{\rm G}$}}
\def\IGa{\relax\hbox{${\rm I}\kern-.18em\Gamma$}}
\def\IH{\relax{\rm I\kern-.18em H}}
\def\II{\relax{\rm I\kern-.18em I}}
\def\IK{\relax{\rm I\kern-.18em K}}
\def\IP{\relax{\rm I\kern-.18em P}}
\def\IQ{\relax\hbox{$\inbar\kern-.3em{\rm Q}$}}

\def\inbar{\,\vrule height1.5ex width.4pt depth0pt}

\font\cmss=cmss10 \font\cmsss=cmss10 at 7pt
\def\IR{\relax{\rm I\kern-.18em R}}

%
%

\def\makeblankbox#1#2{\hbox{\lower\dp0\vbox{\hidehrule{#1}{#2}%
   \kern -#1
   \hbox to \wd0{\hidevrule{#1}{#2}%
      \raise\ht0\vbox to #1{}
      \lower\dp0\vtop to #1{}
      \hfil\hidevrule{#2}{#1}}%
   \kern-#1\hidehrule{#2}{#1}}}%
}%
\def\hidehrule#1#2{\kern-#1\hrule height#1 depth#2 \kern-#2}%
\def\hidevrule#1#2{\kern-#1{\dimen0=#1\advance\dimen0 by #2\vrule
    width\dimen0}\kern-#2}%
\def\openbox{\ht0=1.2mm \dp0=1.2mm \wd0=2.4mm  \raise 2.75pt
\makeblankbox {.25pt} {.25pt}  }

\def\bun#1/#2{\leavevmode
   \kern.1em \raise .5ex \hbox{\the\scriptfont0 #1}%
   \kern-.1em $/$%
   \kern-.15em \lower .25ex \hbox{\the\scriptfont0 #2}%
}

\def\opensquare{\ht0=3.4mm \dp0=3.4mm \wd0=6.8mm  \raise 2.7pt
\makeblankbox {.25pt} {.25pt}  }


\def\sector#1#2{\ {\scriptstyle #1}\hskip 1mm
\mathop{\opensquare}\limits_{\lower 1mm\hbox{$\scriptstyle#2$}}\hskip 1mm}

\def\tsector#1#2{\ {\scriptstyle #1}\hskip 1mm
\mathop{\opensquare}\limits_{\lower 1mm\hbox{$\scriptstyle#2$}}^\sim\hskip 1mm}


\def\inbar{\,\vrule height1.5ex width.4pt depth0pt}

\font\cmss=cmss10 \font\cmsss=cmss10 at 7pt
\def\IR{\relax{\rm I\kern-.18em R}}


\def\frac#1#2{{#1\over#2}}

\def\half{\frac12}

\def\inbar{\,\vrule height1.5ex width.4pt depth0pt}
\def\IC{\relax\hbox{$\inbar\kern-.3em{\rm C}$}}
\def\IR{\relax{\rm I\kern-.18em R}}
\def\IP{\relax{\rm I\kern-.18em P}}

%
%
\catcode`\@=11
\def\slash#1{\mathord{\mathpalette\c@ncel{#1}}}
\overfullrule=0pt

\def\II{{\cal I}}

\def\underrel#1\over#2{\mathrel{\mathop{\kern\z@#1}\limits_{#2}}}

\catcode`\@=12


%

\def\det{{\rm det}}

\def\det{{\rm det}}



\def\frac#1#2{{#1\over#2}}

\def\half{\frac12}

\def\inbar{\,\vrule height1.5ex width.4pt depth0pt}
\def\IC{\relax\hbox{$\inbar\kern-.3em{\rm C}$}}
\def\IR{\relax{\rm I\kern-.18em R}}
\def\IP{\relax{\rm I\kern-.18em P}}

%
%

%
\catcode`\@=11
\def\slash#1{\mathord{\mathpalette\c@ncel{#1}}}
\overfullrule=0pt

\def\II{{\cal I}}

\def\underrel#1\over#2{\mathrel{\mathop{\kern\z@#1}\limits_{#2}}}

\catcode`\@=12


%

\def\det{{\rm det}}

\def\det{{\rm det}}


\lref\ElitzurFH{
  S.~Elitzur, A.~Giveon and D.~Kutasov,
  ``Branes and N = 1 duality in string theory,''
  Phys.\ Lett.\ B {\bf 400}, 269 (1997)
  [arXiv:hep-th/9702014].
}

\lref\ElitzurHC{
  S.~Elitzur, A.~Giveon, D.~Kutasov, E.~Rabinovici and A.~Schwimmer,
  ``Brane dynamics and N = 1 supersymmetric gauge theory,''
  Nucl.\ Phys.\  B {\bf 505}, 202 (1997)
  [arXiv:hep-th/9704104].
}

\lref\AmaritiAM{
   A.~Amariti, L.~Girardello and A.~Mariotti,
   ``Meta-stable $A_n$ quiver gauge theories,''
   arXiv:0706.3151 [hep-th].
}

\lref\AmaritiQU{
   A.~Amariti, L.~Girardello and A.~Mariotti,
   ``On meta-stable SQCD with adjoint matter and gauge mediation,''
   Fortsch.\ Phys.\  {\bf 55}, 627 (2007)
   [arXiv:hep-th/0701121].
}

\lref\AmaritiVK{
   A.~Amariti, L.~Girardello and A.~Mariotti,
   ``Non-supersymmetric meta-stable vacua in SU(N) SQCD with adjoint matter,''
   JHEP {\bf 0612}, 058 (2006)
   [arXiv:hep-th/0608063].
}

\lref\HabaRJ{
  N.~Haba and N.~Maru,
  ``A Simple Model of Direct Gauge Mediation of Metastable Supersymmetry
  Breaking,''
  arXiv:0709.2945 [hep-ph].
}

\lref\GiveonSR{
  A.~Giveon and D.~Kutasov,
  ``Brane dynamics and gauge theory,''
  Rev.\ Mod.\ Phys.\  {\bf 71}, 983 (1999)
  [arXiv:hep-th/9802067].
}

\lref\OoguriBG{
  H.~Ooguri and Y.~Ookouchi,
  ``Meta-stable supersymmetry breaking vacua on intersecting branes,''
  Phys.\ Lett.\ B {\bf 641}, 323 (2006)
  [arXiv:hep-th/0607183].
}

\lref\FrancoHT{
  S.~Franco, I.~Garcia-Etxebarria and A.~M.~Uranga,
  ``Non-supersymmetric meta-stable vacua from brane configurations,''
  arXiv:hep-th/0607218.
}

\lref\BenaRG{
  I.~Bena, E.~Gorbatov, S.~Hellerman, N.~Seiberg and D.~Shih,
  ``A note on (meta)stable brane configurations in MQCD,''
  JHEP {\bf 0611}, 088 (2006)
  [arXiv:hep-th/0608157].
}

\lref\AganagicEX{
  M.~Aganagic, C.~Beem, J.~Seo and C.~Vafa,
  ``Geometrically induced metastability and holography,''
  arXiv:hep-th/0610249.
}

\lref\HeckmanWK{
  J.~J.~Heckman, J.~Seo and C.~Vafa,
  ``Phase Structure of a Brane/Anti-Brane System at Large N,''
  arXiv:hep-th/0702077.
}

\lref\GiveonFK{
  A.~Giveon and D.~Kutasov,
  ``Gauge symmetry and supersymmetry breaking from intersecting branes,''
  Nucl.\ Phys.\  B {\bf 778}, 129 (2007)
  [arXiv:hep-th/0703135].
}

\lref\HananyIE{
  A.~Hanany and E.~Witten,
  ``Type IIB superstrings, BPS monopoles, and three-dimensional gauge
  dynamics,''
  Nucl.\ Phys.\  B {\bf 492}, 152 (1997)
  [arXiv:hep-th/9611230].
}

\lref\DouglasTU{
  M.~R.~Douglas, J.~Shelton and G.~Torroba,
  ``Warping and supersymmetry breaking,''
  arXiv:0704.4001 [hep-th].
}

\lref\MarsanoFE{
  J.~Marsano, K.~Papadodimas and M.~Shigemori,
  ``Nonsupersymmetric brane / antibrane configurations in type IIA and M
  theory,''
  arXiv:0705.0983 [hep-th].
}


\lref\IntriligatorAU{
  K.~A.~Intriligator and N.~Seiberg,
  ``Lectures on supersymmetric gauge theories and electric-magnetic  duality,''
  Nucl.\ Phys.\ Proc.\ Suppl.\  {\bf 45BC}, 1 (1996)
  [arXiv:hep-th/9509066].
}

\lref\PeskinQI{
  M.~E.~Peskin,
  ``Duality in supersymmetric Yang-Mills theory,''
  arXiv:hep-th/9702094.
}

\lref\ShifmanUA{
  M.~A.~Shifman,
  ``Nonperturbative dynamics in supersymmetric gauge theories,''
  Prog.\ Part.\ Nucl.\ Phys.\  {\bf 39}, 1 (1997)
  [arXiv:hep-th/9704114].
}

\lref\StrasslerQG{
  M.~J.~Strassler,
  ``An unorthodox introduction to supersymmetric gauge theory,''
  arXiv:hep-th/0309149.
}

\lref\TerningBQ{
  J.~Terning,
  ``Modern supersymmetry: Dynamics and duality,''
{\it  Oxford, UK: Clarendon (2006) 324 p}
}

\lref\DineZP{
  M.~Dine,
  ``Supersymmetry and string theory: Beyond the standard model,''
{\it  Cambridge, UK: Cambridge Univ. Pr. (2007) 515 p}
}

\lref\IntriligatorCP{
  K.~Intriligator and N.~Seiberg,
  ``Lectures on Supersymmetry Breaking,''
  arXiv:hep-ph/0702069.
}

\lref\IntriligatorDD{
  K.~Intriligator, N.~Seiberg and D.~Shih,
  ``Dynamical SUSY breaking in meta-stable vacua,''
  JHEP {\bf 0604}, 021 (2006)
  [arXiv:hep-th/0602239].
}

\lref\SeibergPQ{
  N.~Seiberg,
  ``Electric - magnetic duality in supersymmetric nonAbelian gauge theories,''
  Nucl.\ Phys.\ B {\bf 435}, 129 (1995)
  [arXiv:hep-th/9411149].
}

\lref\SeibergBZ{
  N.~Seiberg,
  ``Exact Results On The Space Of Vacua Of Four-Dimensional Susy Gauge
  Theories,''
  Phys.\ Rev.\  D {\bf 49}, 6857 (1994)
  [arXiv:hep-th/9402044].
}

\lref\KutasovVE{
  D.~Kutasov,
  ``A Comment on duality in N=1 supersymmetric nonAbelian gauge theories,''
  Phys.\ Lett.\  B {\bf 351}, 230 (1995)
  [arXiv:hep-th/9503086].
}

\lref\KutasovNP{
  D.~Kutasov and A.~Schwimmer,
  ``On duality in supersymmetric Yang-Mills theory,''
  Phys.\ Lett.\  B {\bf 354}, 315 (1995)
  [arXiv:hep-th/9505004].
}

\lref\KutasovSS{
  D.~Kutasov, A.~Schwimmer and N.~Seiberg,
  ``Chiral Rings, Singularity Theory and Electric-Magnetic Duality,''
  Nucl.\ Phys.\  B {\bf 459}, 455 (1996)
  [arXiv:hep-th/9510222].
}

\lref\GiveonEW{
  A.~Giveon and D.~Kutasov,
  ``Stable and Metastable Vacua in Brane Constructions of SQCD,''
  arXiv:0710.1833 [hep-th].
}

\lref\OoguriPJ{
  H.~Ooguri and Y.~Ookouchi,
  ``Landscape of supersymmetry breaking vacua in geometrically realized gauge
  theories,''
  Nucl.\ Phys.\ B {\bf 755}, 239 (2006)
  [arXiv:hep-th/0606061].
}

\lref\KitanoXG{
  R.~Kitano, H.~Ooguri and Y.~Ookouchi,
  ``Direct mediation of meta-stable supersymmetry breaking,''
  arXiv:hep-ph/0612139.
}

\lref\MurayamaYF{
  H.~Murayama and Y.~Nomura,
  ``Gauge mediation simplified,''
  Phys.\ Rev.\ Lett.\  {\bf 98}, 151803 (2007)
  [arXiv:hep-ph/0612186].
}

\lref\NelsonNF{
  A.~E.~Nelson and N.~Seiberg,
  ``R symmetry breaking versus supersymmetry breaking,''
  Nucl.\ Phys.\  B {\bf 416}, 46 (1994)
  [arXiv:hep-ph/9309299].
}

\lref\MurayamaFE{
  H.~Murayama and Y.~Nomura,
  ``Simple scheme for gauge mediation,''
  Phys.\ Rev.\  D {\bf 75}, 095011 (2007)
  [arXiv:hep-ph/0701231].
}

\lref\ArgurioQK{
  R.~Argurio, M.~Bertolini, S.~Franco and S.~Kachru,
  ``Metastable vacua and D-branes at the conifold,''
  JHEP {\bf 0706}, 017 (2007)
  [arXiv:hep-th/0703236].
}

\lref\IntriligatorPY{
  K.~Intriligator, N.~Seiberg and D.~Shih,
  ``Supersymmetry Breaking, R-Symmetry Breaking and Metastable Vacua,''
  JHEP {\bf 0707}, 017 (2007)
  [arXiv:hep-th/0703281].
}

\lref\KawanoRU{
  T.~Kawano, H.~Ooguri and Y.~Ookouchi,
  ``Gauge Mediation in String Theory,''
  Phys.\ Lett.\  B {\bf 652}, 40 (2007)
  [arXiv:0704.1085 [hep-th]].
}

\Title{
} {\vbox{ \centerline{Stable and Metastable Vacua in SQCD}
}}
\medskip
\centerline{\it Amit Giveon${}^{1}$ and David Kutasov${}^{2}$}
\bigskip
\smallskip
\centerline{${}^{1}$Racah Institute of Physics, The Hebrew
University} \centerline{Jerusalem 91904, Israel}
\smallskip
\centerline{${}^2$EFI and Department of Physics, University of
Chicago} \centerline{5640 S. Ellis Av., Chicago, IL 60637, USA }

\bigskip\bigskip\bigskip
\noindent

We study deformations of $N=1$ supersymmetric QCD that exhibit a rich
landscape of supersymmetric and non-supersymmetric vacua.

\vglue .3cm
\bigskip

\Date{9/07}

\bigskip

\newsec{Introduction}

In this paper we study the low energy dynamics of supersymmetric
QCD (SQCD) in the presence of certain F-term deformations. The
starting point of our analysis is an $N=1$ supersymmetric gauge
theory with gauge group $SU(N_c)$ and $N_f>N_c$ flavors of chiral
superfields in the fundamental representation of the gauge group,
$Q_i^\alpha$, $\widetilde Q^i_\alpha$ ($\alpha=1,\cdots, N_c$;
 $i=1,\cdots, N_f$). This theory has a global symmetry
\eqn\symsqcd{SU(N_f)\times SU(N_f)\times U(1)_B\times U(1)_R}
and a non-trivial moduli space of vacua, which has been
extensively studied and is rather well understood; see \eg\
\refs{\IntriligatorAU\PeskinQI\ShifmanUA\StrasslerQG\TerningBQ-\DineZP}
for reviews.

A natural question is what happens when we deform the theory by
adding a general superpotential that preserves a particular
subgroup of the global symmetry \symsqcd, such as the non-chiral
subgroup  $SU(N_f)_{\rm diag}\times U(1)_B$. A class of
superpotentials with this property is \eqn\elsuppot{W_{\rm el}=
\sum_{n=1}^{n_0} {1\over n!}m_n {\rm Tr}M^n} where the meson field
\eqn\defmes{M^i_j=\widetilde Q^i Q_j} is an $N_f\times N_f$ matrix
(the color indices are summed over in \defmes). Terms with $n>1$
in \elsuppot\ are non-renormalizable, which is  reflected in the
fact that the couplings $m_n$ have dimension $[m_n]=3-2n$ at the
free fixed point. One can think of the superpotential \elsuppot\
as providing an effective description below a certain energy
scale.

In the case $m_n=m_1\delta_{n,1}$ \elsuppot\ is a mass term for
$Q$, $\widetilde Q$. It has been known for a long time that the
resulting theory has $N_c$ supersymmetric vacua, in accordance
with the Witten index. More recently, it was found \IntriligatorDD\
that for $N_c<N_f<{3\over2}N_c$ and  small $m_1$ there are metastable
non-supersymmetric ground states as well. Such states might be useful
for describing supersymmetry breaking in nature.

The purpose of this paper is to study more general superpotentials of the 
form \elsuppot.  We will mainly discuss the case where only the two
lowest terms in \elsuppot\ are non-zero, \ie\
\eqn\specwel{W_{\rm el}=m_1{\rm Tr}M+\half m_2 {\rm Tr} M^2}
but will also comment on higher order superpotentials. We will see that
such superpotentials lead generically to a rich landscape of supersymmetric
and non-supersymmetric vacua, and explore some of their properties.
Related works include
\refs{\OoguriPJ\AmaritiVK\KitanoXG\MurayamaYF\AmaritiQU\MurayamaFE\ArgurioQK\IntriligatorPY\KawanoRU\AmaritiAM-\HabaRJ}.

Like in \IntriligatorDD, we will find it useful to utilize the Seiberg dual description
of SQCD, in which the gauge group is $SU(N_f-N_c)$, and the meson \defmes\
becomes a gauge singlet field. We will analyze the dynamics in both the electric
and the magnetic descriptions and compare them.

The deformed SQCD with superpotential \specwel\ has a simple
embedding in string theory, along the lines of
\refs{\HananyIE\ElitzurFH\ElitzurHC\GiveonSR\OoguriBG\FrancoHT\BenaRG\AganagicEX
\HeckmanWK\GiveonFK\DouglasTU-
\MarsanoFE}.
In a companion paper \GiveonEW\ we describe the relevant string construction,
and in particular its connection to the gauge theory results of this paper.
We find that the brane description provides a complementary picture
to the gauge theory one.

The plan of the paper is as follows. In section 2 we construct the supersymmetric ground states
of the theory  \specwel, and  verify that the  electric and magnetic descriptions give
rise to the same vacuum structure once all the relevant quantum effects have been included. In
section 3 we describe metastable states in this theory. In section 4  we discuss our results and
comment on generalizations.

\newsec{Vacuum structure of deformed SQCD}

\subsec{A first look at the vacuum structure}

The $N=1$ supersymmetric Yang-Mills  theory with gauge group
$SU(N_c)$ discussed in the previous section, which we will refer
to as the electric theory, is equivalent in the infrared  \SeibergPQ\
to another gauge theory, which we will refer to as the magnetic
theory. The latter has gauge group $SU(N_f-N_c)$ and
the following set of chiral superfields: $N_f$ fundamentals of the
gauge group, $q^i$, $\tilde q_i$, $i=1,\cdots, N_f$, and a gauge
singlet $M^i_j$ which, as suggested by the notation, is identified
with the gauge invariant meson field \defmes\ in the electric
theory.

The magnetic quarks $q$, $\tilde q$ and meson $M$ are coupled via
the superpotential
\eqn\wmag{W_{\rm mag}={1\over \Lambda}\tilde q_iM^i_j q^j~.}
The scale $\Lambda$ is related to the dynamically
generated scales of the electric and magnetic theories,
$\Lambda_e$, $\Lambda_m$, by the scale matching relation
\IntriligatorAU\
\eqn\scalematch{\Lambda_e^{3N_c-N_f}\Lambda_m^{3(N_f-N_c)-N_f}=
(-)^{N_f-N_c}\Lambda^{N_f}~.}
We are interested in adding to the  electric Lagrangian  the
superpotential
\eqn\suppot{W_{\rm el}= {\alpha\over2}{\rm Tr}(\widetilde Q Q)^2-
m{\rm Tr}(\widetilde Q Q)={\alpha\over2}{\rm Tr}M^2-m {\rm Tr} M}
which has the form \specwel, with $m_1=-m$ and $m_2=\alpha$. In 
the magnetic description this corresponds to deforming \wmag\ to
\eqn\wmagfull{W_{\rm mag}={1\over \Lambda}\tilde q_iM^i_j q^j+
{\alpha\over2}{\rm Tr}M^2-m{\rm Tr}M~.}
Since the superpotential \wmagfull\ is quadratic in $M$, we can integrate
this field out. The resulting superpotential for the magnetic
quarks is given by:
\eqn\effmag{W_{\rm
mag}=-{1\over\alpha\Lambda}\left[{1\over2\Lambda}{\rm Tr}(\tilde q
q)^2-m{\rm Tr}(\tilde q q)\right]~.}
Comparing to \suppot\ we see that the magnetic superpotential has
the same qualitative form as the electric one.

Conversely, one can write the electric superpotential \suppot\ in
a way similar to the magnetic one by integrating in a gauge
singlet field $N$:
\eqn\newelsup{W_{\rm el}=-{1\over\Lambda}\widetilde Q^iN_i^jQ_j-
{\alpha_e\over2} {\rm Tr}N^2+m_e {\rm Tr}N~.}
Requiring that integrating out $N$ leads back to \suppot\ gives
\eqn\constcc{\alpha={1\over\alpha_e\Lambda^2}~,\qquad
m={m_e\over\alpha_e\Lambda}~.}
Comparing \newelsup\ to
\wmagfull\ we see that the two superpotentials have very similar
forms. Note also that we chose the normalization of the field $N$
(which transforms as a singlet under the electric gauge group)
such that it is identified in the infrared with the magnetic quark
bilinear $N=\tilde q q$. This is the dual version of the relation
between the singlet meson $M$ in the magnetic theory and the
bilinear in electric quarks, \defmes.

Although the two forms of the magnetic superpotential, \wmagfull\ and
\effmag, describe the same long distance physics, there is a physical
difference between them, which becomes important for small $\alpha$.
In this limit, the mass of $M$ in \wmagfull\ is small, and at energies above
that mass this field should be included in the dynamics. The description
\effmag\ does not contain this field; it coincides with \wmagfull\ only
at energies well below the mass of $M$. Similar comments apply to the 
electric superpotentials \suppot, \newelsup\ in the limit $\alpha_e\to 0$.

To recapitulate, we see that the analysis of the vacuum structure in the electric and magnetic
descriptions is essentially identical, up to the replacement $N_c\leftrightarrow N_f-N_c$,
$\alpha\leftrightarrow\alpha_e$, $m\leftrightarrow m_e$, etc. We will verify later that
the two descriptions lead to the same vacuum structure, in agreement with Seiberg duality.

At first sight, it actually seems that the electric and magnetic theories have different
vacuum structures. Consider, for example, the classical magnetic superpotential \wmagfull.
To find the supersymmetric vacua we need to solve the F-term constraints
\eqn\fterms{\eqalign{M^i_jq^j=&0~,\cr
                                 \tilde q_iM^i_j=&0~,\cr
                                {1\over\Lambda}\tilde q_iq^j=&m\delta_i^j-\alpha M_i^j~.\cr
                                 }}
{}From \fterms\ we learn that $M$ satisfies the matrix equation
\eqn\constphi{mM=\alpha M^2~.}
On solutions of the equations of motion one can choose $M$ to be diagonal.
Equation \constphi\ implies that its eigenvalues can take only two values, $0$
and $m\over\alpha$. Without loss of generality we can take
\eqn\formphi{M=\left(\matrix{0&0\cr 0&{m\over\alpha}I_{N_f-k}\cr}\right)}
where $k=0,1,2,\cdots, N_f$ and $I_n$ is the $n\times n$ identity matrix.
Plugging \formphi\ into the last line of \fterms\ implies that
\eqn\formqq{\tilde qq=\left(\matrix{m\Lambda I_k&0\cr 0&0\cr}\right)~.}
The rank of the matrix on the left hand side is at most $N_f-N_c$.
Hence, one must have $k\le N_f-N_c$.

Note that the vacua \formphi, \formqq\ go off to infinity in field space
as the deformation parameter $\alpha$ \wmagfull\ goes to zero. In particular,
for $\alpha=0$ supersymmetry is spontaneously broken. This is consistent
with the fact that for $\alpha=0$ the classical superpotential \wmagfull\
has an unbroken $U(1)_R$ symmetry, with $R_q=R_{\tilde q}=0$, $R_M=2$. For
$\alpha\not=0$ this symmetry is explicitly broken, and one expects \NelsonNF\
that supersymmetric vacua exist.

Expanding around the solution \formphi, \formqq\ one finds that the only massless degrees of
freedom are gauge fields and fermions associated with pure $N=1$ SYM corresponding to the
unbroken gauge group $SU(N_f-N_c-k)$.
As mentioned above, quantum mechanically this theory has $N_f-N_c-k$ vacua with a mass gap
(for $N_f-N_c-k\ge2$). Thus, the above analysis implies that the theory with superpotential
\wmagfull\ has the following number of vacua (up to Nambu-Goldstone bosons associated with
broken global symmetries):
\eqn\magvacua{N_{\rm mag}=1+\sum_{k=0}^{N_f-N_c-1}(N_f-N_c-k)
=1+\half(N_f-N_c)(N_f-N_c+1)~.}
The $1$ in \magvacua\ corresponds to the case $k=N_f-N_c$ where the expectation values of the baryon fields
$b=q^{N_f-N_c}$ and $\tilde b=\tilde q^{N_f-N_c}$ are non-zero, which we will refer to as the baryonic branch.

As mentioned above, the analog of the magnetic superpotential \wmagfull\ in the electric theory is \newelsup.
This superpotential clearly leads to the same vacuum structure as \magvacua\ with the replacement $N_f-N_c\to N_c$,
\eqn\nelec{N_{\rm el}=1+\sum_{k=0}^{N_c-1}(N_c-k)=1+\half N_c(N_c+1)~.}
Thus, the electric and magnetic answers are in general different, whereas Seiberg duality implies that
they must agree. The resolution has to do with quantum corrections to the superpotentials of the electric
and magnetic theories. We next turn to a discussion of these corrections, first in the special case
$N_f=N_c+1$, and then in general. We will see that after including quantum effects the number of
supersymmetric vacua is given by
\eqn\nnvac{N_{\rm vac}={\rm max}(N_{\rm el}, N_{\rm mag})}
in both the electric and the magnetic theories.

\subsec{Quantum corrections for $N_f=N_c+1$}

This case is particularly simple, since the magnetic gauge group
is empty. At the same time, according to \nnvac\ the number of
vacua is in this case supposed to be given by the electric result
\nelec, which is larger than the magnetic one \magvacua. The
magnetic quarks $q^i$ are proportional to the electric baryons
$B^i$, and it is convenient to use the latter as the fundamental
degrees of freedom. The magnetic superpotential \wmagfull\ is
known to receive an important correction proportional to $\det M$,
and takes the form \SeibergBZ
\eqn\wwmagg{W_{\rm mag}={1\over\Lambda_e^{2N_f-3}}\left(\widetilde
B_iM^i_j B^j-\det M\right)+ {\alpha\over2}{\rm Tr}M^2-m{\rm Tr}
M~.}
The new term in the superpotential leads to a correction
to the F-term condition on the third line of \fterms, which now
becomes \eqn\eomsusy{ \widetilde B_i B^j-(\det
M)(M^{-1})_i^j+\Lambda_e^{2N_f-3}\left(\alpha
M_i^j-m\delta_i^j\right)=0~.} To analyze the solutions of these
equations we need to distinguish between the cases where the meson
matrix $M$ is regular and singular. Consider first the case where
it is regular, which we will refer to as the mesonic branch. Then,
the first two lines of \fterms\  (with $q^i\propto B^i$) imply
that
\eqn\nobaryon{B^i=\widetilde B_j=0}
while \eomsusy\ takes the form
\eqn\mesbranch{\Lambda_e^{2N_f-3}\left(\alpha M^2-mM\right)=(\det M)I_{N_f}~.}
As before, on solutions of the equations of motion we can diagonalize $M$.
Since the left hand side of \mesbranch\ is proportional to the identity matrix,
the $N_f$ eigenvalues of $M$ must take at most two distinct values, which we
will denote by $x$ and $y$, and are related as follows:
\eqn\sumrule{x+y={m\over\alpha}~.} The vacua split into classes
labeled by an integer $0\le k\le {N_f\over2}$, with $k$ of the
eigenvalues of $M$ equal to $y$, and $N_f-k$ equal to $x$. The
upper bound on $k$ takes into account the freedom of exchanging
$x$ and $y$.

For $k=0$, $M$ is proportional to the identity
matrix, $M=x I_{N_f}$. Plugging this form into \mesbranch\ we find
\eqn\constxko{\Lambda_e^{2N_f-3}(\alpha x-m)=x^{N_f-1}~.}
There are $N_f-1=N_c$ solutions for $x$, and thus
$N_c$ vacua with this form of $M$. For generic $\alpha$, and in
particular when $\alpha$ is sufficiently small, the $N_c$
solutions of \constxko\ are all distinct and non-vanishing, as
expected.

For $k\not=0$ the two distinct eigenvalues of $M$ satisfy
\eqn\formxyk{\eqalign{
\Lambda_e^{2N_f-3}(\alpha x-m)=&x^{N_f-k-1}y^k~,\cr
\Lambda_e^{2N_f-3}(\alpha y-m)=&x^{N_f-k}y^{k-1}~.\cr
}}
Using \sumrule\ we can rewrite both lines of \formxyk\ as
\eqn\eqxkk{-\alpha\Lambda_e^{2N_f-3}=x^{N_f-k-1}y^{k-1}~.}
Since $y$ is linearly related to $x$, \eqxkk\ is a polynomial equation for $x$ of
order $N_f-2$. It has $N_f-2=N_c-1$ distinct solutions with $x,y\not=0$ (again, for generic
$\alpha$). Thus, for odd $N_f$ the number of supersymmetric vacua of the
magnetic theory is given by:
\eqn\nfodd{N_c+\half(N_f-1)(N_c-1)=\half N_c(N_c+1)~.}
For even $N_f$ we have to analyze separately the case $k=N_f/2$ for which the degeneracies of the
two eigenvalues $x$ and $y$ are equal. In this case, due to the symmetry of interchange of $x$
and $y$ we actually have only $\half(N_f-2)=\half(N_c-1)$ distinct solutions of \eqxkk. The total
number of vacua is in this case
\eqn\nfeven{N_c+\half(N_f-2)(N_c-1)+\half(N_f-2)=\half N_c(N_c+1)~.}
We see that for both even and odd $N_f$, the total number of vacua in the magnetic description
\wmagfull\ agrees with \nnvac, except for the contribution of the baryonic branch to which we turn next.

So far we assumed that the meson matrix $M$ is non-degenerate. The
analysis needs to be modified when some of its
eigenvalues vanish. It turns out that the only non-trivial case is
the one in which exactly one eigenvalue of $M$ vanishes. Thus, we
take
\eqn\diagmm{M={\rm diag}(M_1,M_2,\cdots, M_{N_f})}
and consider the limit $M_1\to 0$ with the rest of the $M_j$ remaining finite. The first
two lines of \fterms\ imply that the only non-zero components of $B$, $\widetilde B$ are
$B^1$, $\widetilde B_1$. Eq. \eomsusy\ leads to:
\eqn\barbranch{\eqalign{M_j=&{m\over\alpha}, \;\;j=2,\cdots,N_f~,\cr
B^1\widetilde B_1=&\left({m\over\alpha}\right)^{N_f-1}+m\Lambda_e^{2N_f-3}~.\cr
}}
Including this extra vacuum, which corresponds to the baryonic branch in the
classical analysis, brings the total number of vacua into agreement with \nnvac.

If $k>1$ eigenvalues of $M$ go to zero, \eomsusy\ has no
solutions. Indeed, in that case the first two lines of \fterms\
imply that the non-zero components of $B$, $\widetilde B$
lie in the degenerate, $k$ dimensional, subspace. In that
subspace, \eomsusy\ implies that \eqn\norank{B^i\widetilde
B_j\propto\delta^i_j~,} which is impossible to satisfy since the
left hand side has rank 1 while the right hand side has rank $k>1$.

To summarize, for $N_f=N_c+1$ we conclude that the number of vacua of the magnetic theory is given by
\nnvac. The term that goes like $\det M$ in the magnetic superpotential \wwmagg\ is crucial for the
analysis. The analog of this term for generic $N_f>N_c+1$ comes from quantum effects in the magnetic
theory. We next turn to a discussion of these effects.

\subsec{Quantum corrections for $N_f>N_c+1$}

We would like to extend the analysis of the previous subsection
to general $N_f$. Consider a vacuum in which the expectation value
of the meson field $M$ has $r$ vanishing eigenvalues and $N_f-r$
non-vanishing ones. As mentioned in subsection 2.1, the F-term
constraint on the third line of \fterms\ implies that $r\le
N_f-N_c$. Therefore, at least $N_c$ eigenvalues of $M$ must be
non-zero. In particular, we can bring it to the form
\eqn\redflav{M=\left(\matrix{\widehat{M}&0\cr
                             0&M_0\cr
                             }\right)}
where $M_0$ is a non-degenerate $(N_c-1)\times (N_c-1)$ matrix, and
$\widehat M$ is an $(N_f-N_c+1)\times(N_f-N_c+1)$ one, which may or may not
be degenerate (but whose rank is at least one).

Since $M_0$ is non-degenerate, the flavors of quarks $q$, $\tilde q$ that couple
to it via \wmag\ are massive and can be integrated out at low energies. This leads
to an $SU(N_f-N_c)$ gauge theory with $N_f-N_c+1$ flavors $q^a$, $\tilde q_{\tilde a}$
($a,\tilde a=1,\cdots, N_f-N_c+1$), whose scale is given by
\eqn\lowscale{\Lambda_L^{2(N_f-N_c)-1}={\det M_0\over\Lambda^{N_c-1}}
\Lambda_m^{3(N_f-N_c)-N_f}~.}
This theory is of the sort discussed in the previous subsection. It can be described in terms of the
gauge invariant observables
\eqn\gaugeinv{N^a_{\tilde a}=\tilde q_{\tilde a}\cdot q^a~,\qquad
b_a=q^{N_f-N_c}~,\qquad
\tilde b^{\tilde a}= \tilde q^{N_f-N_c}~.}
The full magnetic superpotential has the form
\eqn\fullmagsup{W_{\rm mag}={\rm Tr}\left({1\over\Lambda}\widehat M N+{\alpha\over2}\widehat M^2
-m\widehat M\right)+{\rm Tr}\left({\alpha\over2}M_0^2-m M_0\right)
+{1\over\Lambda_L^{2(N_f-N_c)-1}}\left(\tilde b\cdot N \cdot b-\det N\right)~.}
The field $\widehat M$ appears quadratically in \fullmagsup\ and thus can be integrated out.
The equation of motion of $\widehat{M}$ sets it to
\eqn\valhatm{\widehat M=-{1\over\alpha\Lambda}(N-m\Lambda I_{N_f-N_c+1})~.}
Plugging this into \fullmagsup\ (and dropping a constant contribution to the superpotential)
leads to
\eqn\nnsup{\eqalign{W_{\rm mag}=-&{1\over\alpha\Lambda}{\rm Tr}\left({1\over2\Lambda}N^2-m N\right)
+{\rm Tr}\left({\alpha\over2}M_0^2-mM_0\right)\cr
+&{\Lambda^{N_c-1}\over\Lambda_m^{3(N_f-N_c)-N_f}\det M_0}\left(\tilde b \cdot N \cdot b-\det N\right)~.\cr}}
The part of the superpotential that depends on $N$, $b$, $\tilde b$ is very similar to
the one analyzed in the previous subsection, \wwmagg. The new element is the dependence on
$M_0$ which needs to be taken into account.

As before, the F-term equations of motion of $N$, $b$, $\tilde b$ have two kinds of solutions:
one in which $N$ is non-degenerate, and another in which it has exactly one vanishing eigenvalue.

Consider first the mesonic branch, where $N$ is non-degenerate. In this case
it is convenient to go back to \fullmagsup\ and integrate out $N$,
$b$, $\tilde b$. This amounts to setting $b=\tilde b=0$ in
\fullmagsup\ and replacing $N$ by the solution of its equation of
motion, \eqn\nneom{\widehat{M} N={\Lambda^{N_c}\over
\Lambda_m^{3(N_f-N_c)-N_f}}{\det N\over\det M_0} I_{N_f-N_c+1}~.}
Solving for $N$ and substituting in \fullmagsup\ (using
\scalematch) we get \eqn\fullsupmm{W_{\rm mag}={\rm
Tr}\left({\alpha\over2}M^2-mM\right)-(N_f-N_c) \left(\det
M\over\Lambda_e^{3N_c-N_f}\right)^{1\over N_f-N_c}} where $M$ is
the full meson matrix \redflav, which is non-degenerate in this
case. Of course, the determinant term in \fullsupmm\ is nothing
but the well known non-perturbative superpotential
\refs{\IntriligatorAU\PeskinQI\ShifmanUA\StrasslerQG\TerningBQ-\DineZP}
for the meson field in SQCD. We could have gotten it directly from
\wmagfull\ by assuming that $M$ is a non-degenerate matrix and
integrating out the massive quarks $q$, $\tilde q$. The fact that
we got it with the correct coefficient is a check on the algebra.

We can now generalize the discussion of the previous subsection and look for (non-singular)
solutions of the F-term equations corresponding to the superpotential \fullsupmm:
\eqn\nonsing{\alpha M^2-mM=\left(\det M\over\Lambda_e^{3N_c-N_f}\right)^{1\over N_f-N_c}I_{N_f}~.}
Again, the matrix $M$ has only two distinct eigenvalues $x$,  $y$ satisfying the relation \sumrule,
and vacua are labeled by an integer $k$ which keeps track of the number of times the eigenvalue
$y$ (say) appears.

Considerations similar to those that led to \eqxkk\ give
\eqn\relxygen{x^{N_c-k}y^{k-N_f+N_c}=(-\alpha)^{N_f-N_c}\Lambda_e^{3N_c-N_f}~.}
Together with \sumrule\ this gives a polynomial equation for $x$, whose order depends
on $N_f, N_c$ and $k$.

Consider first the case $N_f\le 2N_c$. For $k\le N_f-N_c$, \relxygen\ is a polynomial of degree
$N_c-k$. Thus it has $N_c-k$ solutions. For $N_f-N_c\le k\leq {N_f\over2}$ the degree of the
polynomial and the number of solutions are given by $2N_c-N_f$. Summing over $k$ one finds
that the number of vacua (up to global symmetries) is $\half N_c(N_c+1)$, as before \nfodd.

For $N_f\ge 2N_c$ the picture is slightly different. For $0\le k
\le N_c$ one finds $N_f-N_c-k$ solutions, and for $N_c\le k \le
{N_f\over2}$, $N_f-2N_c$ solutions. The total in this case is
$\half(N_f-N_c)(N_f-N_c+1)$, again in agreement with \nnvac.

It is not surprising that for $N_f>2N_c$ the number of vacua agrees with the magnetic answer,
while for $N_f<2N_c$ it does not. The non-perturbative superpotential in \fullsupmm\ gives a
correction to the classical superpotential for $M$, \wmagfull, that goes like $M^{N_f\over N_f-N_c}$.
For $N_f>2N_c$ this correction is subleading at large $M$ relative to the leading, $M^2$, term,
and one does not expect it to change the number of vacua. On the other hand, for $N_f<2N_c$
it changes the behavior of the potential at infinity, and it is natural that the number of vacua changes.

So far we assumed that the matrix $N$ \gaugeinv\ is regular. As we saw in the previous subsection,
the only other case we need to consider is the baryonic branch, in which the rank of $N$ is $N_f-N_c$
(\ie\ one of the eigenvalues of $N$ goes to zero). We can choose this eigenvalue
to be $N_1^1$. The F-term equations of the superpotential \fullmagsup\ take in this case the form:
\eqn\ftermbar{\eqalign{
\widehat{M}^1_1=&{\Lambda\over\Lambda_L^{2(N_f-N_c)-1}}\left[[\det N]_{11}-b_1\tilde b^1\right]
={m\over\alpha}~;\cr
\widehat{M}_i^j=&0;\;\;N_i^j=m\Lambda\delta_i^j,\;\;\;      i,j>1~;\cr
M_0=&{m\over\alpha} I_{N_c-1}~,\cr
}}
where $[\det N]_{11}$ is the determinant of $N$ with the first row and column discarded (\ie\
the $(11)$ minor of the matrix $N$).

Thus, the meson matrix $M$ takes the form
\eqn\mesonmat{M=\left(\matrix{M^{(1)}&0\cr
0&0\cr}\right)}
where $M^{(1)}$ is proportional to the $N_c\times N_c$ identity matrix,
\eqn\mone{M^{(1)}={m\over\alpha}I_{N_c}~.}
The $(N_f-N_c+1)\times (N_f-N_c+1)$ matrix $N$ has a block proportional to the $(N_f-N_c)$
dimensional identity matrix \ftermbar. Overall, we find that the number of vacua in the
magnetic gauge theory with the superpotential \wmagfull\ agrees precisely with \nnvac.

\subsec{Matching the electric and magnetic descriptions}

In the previous two subsections we performed a detailed analysis of the supersymmetric vacua
of the magnetic gauge theory \wmagfull, and in particular reproduced \nnvac\ in that description.
It is interesting to check this result in the electric description. In fact, this does not require any
additional work. As mentioned above, the description of the electric gauge theory via the
superpotential \newelsup\ is identical to the magnetic one \wmagfull\ with the substitutions
\eqn\substit{\eqalign{
&N_c\to  N_f-N_c~,\cr
&\Lambda_m \to  \Lambda_e~,\cr
&\Lambda\to -\Lambda~,\cr
&(m,\alpha)\to  -(m_e,\alpha_e)~,\cr
&(q, \tilde q)\to  (Q,\widetilde Q)~,\cr
&M\to  N~.\cr
}}
As a check, note that the transformation on the first three lines of \substit\ is a symmetry of the scale
matching condition \scalematch.

To see how the vacua we found in the previous subsections arise in the electric description \newelsup,
consider for example the case $N_c<N_f<2N_c$ (the regime $N_f>2N_c$ is very similar). In the magnetic
description we found vacua in which the meson matrix $M$ \defmes\ takes the form (up to global symmetries)
\eqn\formmmm{M={\rm diag}(x^{N_f-k}, y^k)}
with $x$ and $y$ satisfying the relations \sumrule, \relxygen. The number of solutions as a function of $k$
is $N_c-k$ for $0\le k\le N_f-N_c$, and $2N_c-N_f$ for $N_f-N_c\le k\le {N_f\over2}$.

To analyze this case in the electric variables we need to use the
map \substit. Due to the transformation of the number of colors,
the electric theory is actually in the opposite regime of the
analysis of subsection 2.3, $N_f>2\widetilde N_c=2(N_f-N_c)$.
According to that analysis, the electric meson
matrix $N$ takes in this case the form \eqn\formeee{N={\rm
diag}(x_e^{N_f-k}, y_e^k)} where now for $0\le k\le \widetilde
N_c=N_f-N_c$ there are $N_f-\widetilde N_c-k=N_c-k$ solutions, and
for $N_f-N_c=\widetilde N_c\le k\le {N_f\over2}$ there are
$N_f-2\widetilde N_c=2N_c-N_f$ solutions. Comparing to the
magnetic analysis, we see that the degeneracies are exactly the
same for all $k$, and we should identify the electric and magnetic
vacua for each value of $k$ separately.

We next show that the detailed form of the meson matrix one finds in
the electric and magnetic descriptions is indeed the same for each $k$.
To facilitate the comparison, we need to translate the results of the electric
analysis, which give the auxiliary gauge singlet meson matrix $N$ to those
for the meson matrix $M$ \defmes. The equation of motion of $N$ arising
from the superpotential \newelsup\ gives the relation between the two:
\eqn\relmmnn{{M\over\Lambda}=m_e I_{N_f}-\alpha_e N~.}
Plugging \formmmm, \formeee\ into \relmmnn\ we find the following
relations between the eigenvalues:
\eqn\releigen{\eqalign{x=&\Lambda(m_e-\alpha_e x_e)~,\cr
                       y=&\Lambda(m_e-\alpha_e y_e)~.\cr
}}
These relations can be simplified by recalling that $x_e$ and $y_e$
satisfy an analog of \sumrule\ obtained by making the replacements
\substit,
\eqn\elrelxy{x_e+y_e={m_e\over\alpha_e}}
using which we can rewrite \releigen\ as
\eqn\simpreleigen{\eqalign{x=&\alpha_e\Lambda y_e~,\cr
                           y=&\alpha_e\Lambda x_e~.\cr
}}
The non-trivial check is that the polynomial equation satisfied by $x$,
$y$, \relxygen, and the corresponding equation for $x_e$, $y_e$,
\eqn\relxeye{x_e^{N_f-N_c-k}y_e^{k-N_c}=(+\alpha_e)^{N_c}\Lambda_m^{3(N_f-N_c)-N_f}}
are compatible for all $k$. Plugging \simpreleigen\ into
\relxygen\ and using the scale matching condition \scalematch\ as
well as \constcc\ one finds that this is indeed the case.

\subsec{The classical limit}

In the previous subsections we found that the number of supersymmetric vacua of the deformed SQCD
system described by \suppot\ -- \newelsup\ is given by \nnvac. For $N_f\ge2N_c$ it agrees with the
classical analysis of the magnetic theory, while in the electric description some of the vacua exist
only in the quantum theory and go off to infinity in the classical limit. For $N_f\le 2N_c$ it is
the other way around.

To see how this happens in detail, we can take the classical limit
of our general results. Consider, for example, the limit in which
the magnetic theory becomes classical,\foot{The analysis
of the classical limit in the electric theory is very similar.}
$\Lambda_m^{3\widetilde N_c-N_f}\to0$. In this limit (which is
equivalent via \scalematch\
to $\Lambda_e^{3N_c-N_f}\to\infty$) the quantum corrections to the
classical superpotential vanish (see \eg\ \fullsupmm). To see what
happens to the quantum vacua, we need to analyze the behavior of
the eigenvalues $(x,y)$ of subsection 2.3 in this limit.

Consider, for example, the case $N_f\le2N_c$. The equations for the eigenvalues, \sumrule, \relxygen\ take
different forms for different $k$. For $0\le k\le N_f-N_c$ one has
\eqn\agaa{x^{N_c-k}=(-\alpha)^{N_f-N_c}\Lambda_e^{3N_c-N_f}\left({m\over\alpha}-x\right)^{N_f-N_c-k}~.}
This is a polynomial equation of degree $N_c-k$. The solutions exhibit two
types of behavior in the classical limit. There are $N_f-N_c-k$ solutions of the form
\eqn\classsol{x={m\over\alpha}-\epsilon~,\qquad y=\epsilon~,}
with
\eqn\formeps{\epsilon^{N_f-N_c-k}\simeq(-\alpha)^{N_c-N_f}\left(m\over\alpha\right)^{N_c-k}\Lambda_e^{N_f-3N_c}~.}
Note that $\epsilon\to 0$ in the classical limit. Thus, these solutions are small deformations of the
classical solutions found in subsection 2.1.

In addition, \agaa\ has $2N_c-N_f$ solutions in which
\eqn\largex{x^{2N_c-N_f}\simeq(-)^k\alpha^{N_f-N_c}\Lambda_e^{3N_c-N_f}~.}
These solutions go to $x=\infty$ in the classical limit.

{}For $N_f-N_c\leq k\leq{N_f\over 2}$ \agaa\ is a polynomial
equation of degree $2N_c-N_f$. Its solutions have the form
\largex, so they too go to infinity in the classical limit. The
baryonic vacuum \ftermbar\ is a small deformation of the classical
one.

{}For $N_f\geq 2N_c$ one can perform a similar analysis and verify
that all the solutions of  \relxygen\ are small deformations of
the classical ones, as expected.

\newsec{Metastable vacua}

In the previous section we discussed the supersymmetric ground
states of deformed SQCD \specwel. In this section, we will see
that this theory has many non-supersymmetric metastable ground 
states as well. As in \IntriligatorDD, we will restrict attention to the 
regime $N_f< {3\over2}N_c$, where the magnetic theory is free in 
the infrared, and can be thought of as the effective low energy
description of the asymptotically free electric gauge theory.
A similar analysis can be performed in the free electric
phase.\foot{Or, more generally, whenever one of the descriptions
is weakly coupled at the energy scale associated with the metastable vacua.}

The Kahler potential of the singlet meson superfield $M$ in
\wmagfull\ in the free magnetic phase is expected to take the form
\eqn\kahlmm{K={1\over a|\Lambda_e|^2}M^\dagger M+\cdots}
near the origin $M=0$. The positive real constant $a$ is not easy to calculate (see
\IntriligatorDD\ for further discussion). It is convenient to
define a superfield $\Phi$ via
\eqn\relmphi{M=\sqrt{a}\Lambda_e\Phi}
such that the Kahler potential for $\Phi$ and the magnetic quarks is canonical
near the origin of field space,
\eqn\cankah{K={\rm Tr}q^\dagger q+{\rm Tr}\tilde q^\dagger\tilde q
+{\rm Tr}\Phi^\dagger\Phi+\cdots~.}
Corrections to the Kahler potential \cankah\ are due to physics at or above the
scale $\Lambda_m$ where the magnetic gauge theory breaks down and is replaced by
the asymptotically free dual electric theory discussed above.
The leading corrections are expected to be quartic in the fields and suppressed by
two powers of $\Lambda_m$ \IntriligatorDD,
\eqn\corrkah{\delta K\sim {1\over|\Lambda_m|^2}{\rm Tr}(\Phi^\dagger\Phi)^2+\cdots~.}
In order to be able to ignore them we will require the expectation values of the
fields $q$, $\tilde q$, $\Phi$ to be much smaller than $|\Lambda_m|$.

The magnetic superpotential \wmagfull\ is given by
\eqn\wwmm{W_{\rm mag}=h\tilde q_i\Phi^i_j q^j- {\rm Tr}\left(h\mu^2\Phi-\half
h^2\mu_\phi\Phi^2\right)= {1\over\Lambda}\tilde q_iM^i_j q^j+ {\rm
Tr} \left(\half \alpha M^2-mM\right)~.}
By equating the two expressions in \wwmm\  we can relate the
parameters as follows:
\eqn\relpar{h=\sqrt{a}{\Lambda_e\over\Lambda}~, \qquad
\mu^2=m\Lambda~,\qquad \mu_\phi=\alpha\Lambda^2~.}
The classical supersymmetric vacua \formphi, \formqq\ take in these
variables the form
\eqn\formmm{h\Phi=\left(\matrix{0&0\cr 0&{\mu^2\over\mu_\phi}I_{N_f-k}\cr}\right)~,}
\eqn\formqqqq{\tilde qq=\left(\matrix{\mu^2 I_k&0\cr 0&0\cr}\right)~.}
To construct the metastable states, it is convenient to further split the
$(N_f-k)\times (N_f-k)$ block at the lower right corner of
\formmm, \formqqqq\ into blocks of size $n$ and $N_f-k-n$ as
follows:
\eqn\metaone{h\Phi=\left(\matrix{0&0&0\cr 0&h\Phi_n&0\cr
0&0&{\mu^2\over\mu_\phi}I_{N_f-k-n}\cr}\right)~,}
and
\eqn\metatwo{\tilde qq=\left(\matrix{\mu^2 I_k&0&0\cr
0&\tilde\varphi\varphi&0\cr 0&0&0}\right)~.}
$\varphi$ and $\tilde\varphi$ are $n\times (N_f-N_c-k)$ dimensional matrices.
They correspond to $n$ flavors of  fundamentals of the gauge group $SU(N_f-N_c-k)$
which is unbroken by the non-zero expectation value of $q$, $\tilde q$ in \metatwo.
$\Phi_n$ and $\tilde\varphi\varphi$ are $n\times n$ matrices. The supersymmetric
ground state \formmm, \formqqqq\ corresponds to $h\Phi_n={\mu^2\over\mu_\phi}I_n$,
$\varphi=\tilde\varphi=0$.

As we will see next, there are metastable vacua near the origin as well. We will
restrict to the regime
\eqn\mumuphi{\Lambda_m\gg\mu\gg\mu_\phi}
where the first inequality is as in \IntriligatorDD, and the second implies that
the term proportional to $\mu_\phi$ is a small perturbation of the superpotential
considered in \IntriligatorDD. Since in general the expectation value of $\Phi$
\metaone\ can be large in the regime \mumuphi, it is convenient to discuss separately
the cases $n=N_f-k$ and $n<N_f-k$.

\subsec{Metastable vacua with $n=N_f-k$}

In this case, the dynamics of the fields $\Phi_n$, $\varphi$, $\tilde\varphi$
near the origin of field space is obtained from the underlying $SU(N_f-N_c)$
gauge theory by giving an expectation value\foot{In the regime  \mumuphi, the
expectation value of $q$ is such that the corrections to the Kahler potential
discussed around eq. \corrkah\ can be neglected, as in \IntriligatorDD.} $\mu$
to $k$ flavors of magnetic
quarks $q$, $\tilde q$, \metatwo. The low energy theory is an $SU(N_f-N_c-k)$
gauge theory with $N_f-k=n$ light flavors, which is infrared free in the
regime $N_f<{3\over 2}N_c$. Its scale, $\Lambda_l$, is related
to that of the underlying theory, $\Lambda_m$, via the scale matching
relation
\eqn\higgssm{\Lambda_m^{3(N_f-N_c)-N_f}=\mu^{2k}\Lambda_l^{3(N_f-N_c-k)-n}~.}
A useful way of writing \higgssm\ is
\eqn\anway{\left(\mu\over\Lambda_m\right)^{2k}=\left(\Lambda_m\over\Lambda_l\right)^{3(N_f-N_c-k)-n}~.}
In the regime \mumuphi, the left hand side of \anway\ is very small. Since the
power on the right hand side is negative, we conclude that
$\Lambda_m\gg\Lambda_l$. Using the fact that \higgssm\ also implies that
\eqn\thirdway{\left(\mu\over\Lambda_l\right)^{2k}=\left(\Lambda_m\over\Lambda_l\right)^{3(N_f-N_c)-N_f}}
we conclude that the hierarchy of scales is
\eqn\hiersc{\mu\ll\Lambda_l\ll\Lambda_m~.}
Thus, we see that as long as $\mu\ll\Lambda_m$, the gauge dynamics is weakly coupled
for all energies well below $\Lambda_m$. At energies above $\mu$, the full magnetic
gauge group is restored. At an energy of order $\mu$ the theory crosses over to the
one with gauge group $SU(N_f-N_c-k)$, which is also infrared free and due to \hiersc\
is weakly coupled. Therefore, we can mostly neglect the gauge dynamics in what follows.

The potential for $\Phi_n$, $\varphi$, $\tilde\varphi$ near the origin of field space
has two relevant contributions. One is the tree level potential that follows from
\cankah, \wwmm. The second is the one loop potential, which to leading order in $h$,
$\mu_\phi\over\mu$ is identical to that computed in \IntriligatorDD, and is given by
\eqn\oneloop{V_{\rm 1-loop}=b|h^2\mu|^2{\rm Tr}\Phi_n^\dagger\Phi_n}
with $b$ a numerical constant,
\eqn\bform{b={\ln4-1\over8\pi^2}(N_f-N_c)~.}
The full one loop potential for $\Phi_n$, $\varphi$, $\tilde\varphi$
has the form
\eqn\vfull{{V\over |h|^2}=|\Phi_n\varphi|^2+|\tilde\varphi\Phi_n|^2+
|\tilde\varphi\varphi-\mu^2I_n+h\mu_\phi\Phi_n|^2+b|h\mu|^2{\rm
Tr}\Phi_n^\dagger\Phi_n~.}
We are looking for a minimum of the
potential at $\varphi=\tilde\varphi=0$. Differentiating \vfull\
w.r.t. $\Phi_n$ we find that if there is one, it is located at
\eqn\hhphi{h\Phi_n={\mu^2\mu_\phi^*\over|\mu_\phi|^2+b|\mu|^2}I_n\simeq
{\mu^2\mu_\phi^*\over b|\mu|^2}I_n~,}
and its vacuum energy is given to leading order in $h$, $\mu_\phi\over\mu$ by
\eqn\vacen{V\simeq n|h\mu^2|^2~.}
Expanding around this solution one finds that the mass matrix of
$\varphi$, $\tilde\varphi$ has eigenvalues
\eqn\massmatrix{m_\pm^2={|\mu|^4\over (|\mu_\phi|^2+b|\mu|^2)^2}
\left[|\mu_\phi|^2\pm b|h|^2(|\mu_\phi|^2+b|\mu|^2)\right]\simeq
{1\over b^2}\left(|\mu_\phi|^2\pm |bh\mu|^2\right)~.}
Equation \massmatrix\ implies that for sufficiently small $\mu_\phi$ some
of the fluctuations of $\varphi$, $\tilde\varphi$ are tachyonic. To avoid
tachyons one must have
\eqn\boundmuphi{\left|\mu_\phi\over\mu\right|^2>{|bh|^2\over 1-b|h|^2}
\simeq |bh|^2~.}
The condition \boundmuphi\ is compatible with $|h|$, $\left|\mu_\phi\over\mu\right|$
being arbitrarily small. If it is satisfied, the quarks $\varphi$, $\tilde\varphi$ are
massive, and the vacuum \hhphi\ is locally stable.

{}For $k<N_f-N_c-1$, this vacuum contains an unbroken $SU(N_f-N_c-k)$ gauge
theory, which is weakly coupled at energies above the scale of the masses of
$\varphi$, $\tilde\varphi$ \massmatrix. For energies well below these masses
this gauge theory confines and has, as before, $N_f-N_c-k$ vacua. For
$k=N_f-N_c-1$ there are no unbroken gauge fields and the theory is weakly
coupled at long distances.

{}For $k=N_f-N_c$ (the largest value it can take), the fields $\varphi$,
$\tilde\varphi$ in \metatwo\ do not exist. Hence, in this case the constraint
\boundmuphi\ is absent and the resulting metastable vacuum is present for
arbitrarily small $\mu_\phi$. In the limit $\mu_\phi\to 0$ it corresponds to
the metastable state discussed in \IntriligatorDD. Even when \boundmuphi\ is
satisfied, the vacuum with $k=N_f-N_c$ is more long-lived than those with
$k<N_f-N_c$ since the mode of instability towards condensation of $\varphi$,
$\tilde\varphi$ is absent in this case.

\subsec{Metastable vacua with $n<N_f-k$}

In order to trust an analysis based on the Kahler potential \cankah\ we have
to demand that all components of $\Phi$ \metaone\ are much smaller than $\Lambda_m$.
In the present case this implies
\eqn\constmuphi{{\mu^2\over\mu_\phi}\ll h\Lambda_m~.}
A useful way of thinking about \constmuphi\ is as the requirement
\eqn\newway{{\mu\over\Lambda_m}\ll h{\mu_\phi\over\mu}~,}
which is compatible with all three couplings $\mu\over\Lambda_m$, $h$ and
$\mu_\phi\over\mu$ being small. Furthermore, this requirement is very natural
in the brane realization of the theory, \GiveonEW.

Like in the previous subsection, we would like the gauge dynamics to be weak
near the origin of the field space of $\varphi$, $\tilde\varphi$, $\Phi_n$,
\metaone, \metatwo. The dynamics of these fields is obtained from the magnetic
$SU(N_f-N_c)$ gauge theory by giving masses $\mu^2\over\mu_\phi$ to $N_f-k-n$
flavors (see \metaone) and vacuum expectation values $\mu$ to $k$ flavors (see
\metatwo). In the regime \mumuphi, the masses are much larger than the
expectation values. Thus we can analyze the reduction from the
$SU(N_f-N_c)$ magnetic gauge theory to the low energy $SU(N_f-N_c-k)$ one in two
steps, by first incorporating the masses, and then the expectation values. We would
like the theory to remain weakly coupled throughout this process.

In the first step, we go from an $SU(N_f-N_c)$ SYM theory with $N_f$ flavors
and scale $\Lambda_m$ to one with the same gauge group, $n+k$ flavors and scale
$\Lambda_1$. The scale matching relation between the two theories is
\eqn\scalemass{\Lambda_m^{3(N_f-N_c)-N_f}
\left(\mu^2\over\mu_\phi\right)^{N_f-k-n}=\Lambda_1^{3(N_f-N_c)-(k+n)}~.}
The fact that the magnetic gauge theory is not asymptotically free is reflected
in the power of $\Lambda_m$ on the left hand side of \scalemass\ being
negative. The theory with $n+k$ flavors can be either asymptotically free or not,
which is reflected in the fact that the power of $\Lambda_1$ on the right hand side
can be positive or negative.

Consider first the case\foot{In the free magnetic phase $N_f>3(N_f-N_c)$
there are solutions to this constraint. For large generic $N_f$, $N_c$, the
number of such solutions is large as well.}
\eqn\nonasfreekn{k+n>3(N_f-N_c)}
in which it is not asymptotically free. Rewriting \scalemass\ in the form
\eqn\ratioone{\left({\mu^2\over\mu_\phi}\over\Lambda_m\right)^{N_f-k-n}=
\left({\Lambda_1\over\Lambda_m}\right)^{3(N_f-N_c)-k-n}}
and using the fact that the left hand side is very small due to \constmuphi,
we conclude that
\eqn\nonas{\Lambda_1\gg\Lambda_m~.}
Hence the $SU(N_f-N_c)$ gauge dynamics is weakly coupled both for energies
larger than $\mu^2\over\mu_\phi$, for which the number of flavors is $N_f$
and for lower energies, where it is $n+k$.

In the second step of the reduction we give an expectation value $\mu$ to $k$
flavors of quarks and go down to an $SU(N_f-N_c-k)$ gauge theory with $n$ light
flavors and scale $\Lambda_l$. The scale matching relation is (compare to \higgssm)
\eqn\scaletwo{\Lambda_1^{3(N_f-N_c)-(k+n)}=\mu^{2k}\Lambda_l^{3(N_f-N_c-k)-n}~.}
In the regime \nonasfreekn\ the power of $\Lambda_l$ in \scaletwo\ is negative.
Thus, the low energy $SU(N_f-N_c-k)$ gauge theory with $n$ flavors is weakly
coupled below the scale $\Lambda_l$. Using the fact that $\mu\ll\Lambda_1$,
one can see from \scaletwo\ that $\Lambda_l$ is in the range
\eqn\mmuull{\mu\ll\Lambda_l\ll\Lambda_1~.}
We see that the theory is weakly coupled at all scales. At energies much above $\mu$ it is governed
by the intermediate theory with scale $\Lambda_1$, which as we saw before is weakly coupled.
At the energy $\mu$ it crosses over to the low energy theory with the scale $\Lambda_l$, which
due to \mmuull\ is weakly coupled as well. Consequently, the gauge dynamics can be neglected,
as in the previous subsection, and one can proceed as there, with similar conclusions.

To summarize, we find that deformed SQCD in the regime  \mumuphi, \newway\ in coupling space
has metastable vacua of the form \metaone, \metatwo, \hhphi, with $k$ and $n$ satisfying the
constraint \nonasfreekn. The states with $k<N_f-N_c$  further require \boundmuphi\ for their
existence. Those with $k=N_f-N_c$ exist throughout this regime.

The above analysis can be repeated for the case where the constraint \nonasfreekn\ is not satisfied.
One finds that there are regions in parameter space in which the gauge dynamics is weakly coupled
and metastable states exist. We will not describe the details here.

Finally, it is interesting to ask what happens to the states with $k=N_f-N_c$, $n<N_c$ as $\mu_\phi\to 0$.
As $\mu_\phi$ decreases, at some point we leave the weak coupling regime \constmuphi\ and the
analysis presented above becomes unreliable. The limit $\mu_\phi\to 0$ involves strongly coupled
dynamics. The brane construction of \GiveonEW\ seems to suggest that these vacua survive in the limit,
and thus in massive SQCD there are additional metastable states to those studied in \IntriligatorDD.
More work is required to resolve this issue conclusively.

\newsec{Discussion}

In this paper we discussed supersymmetric QCD in the presence of
the superpotential \suppot. We described the supersymmetric vacua
of the theory, and generalized the discussion of metastable
supersymmetry breaking states in \IntriligatorDD\ to this case. We
saw that by tuning the parameters of the theory one can make the
non-supersymmetric states arbitrarily long-lived. In other regions
of parameter space, which can be studied using weakly coupled
field theory, many of these states become unstable and disappear.

The mechanism for the emergence of metastable states in the theory
with $\alpha\not=0$ \suppot\ is slightly different than that in the theory
with $\alpha=0$ studied in \IntriligatorDD. There, the classical theory 
spontaneously broke supersymmetry and had a pseudo-moduli space of 
non-supersymmetric states. One loop effects lifted the pseudo-moduli 
space and replaced it by an isolated metastable state. In our case, classically
there are supersymmetric ground states and {\it no} non-supersymmetric
metastable states. The latter are due to a competition between the classical 
and one loop contributions to the potential. The fact that one can ignore higher
loop contributions to the potential in studying these states is analogous to what 
happens in the $\epsilon$ expansion in quantum field theory.

The theory with quartic superpotential in \suppot\ requires a UV
completion. One possibility is to introduce the singlet meson $N$,
and rewrite it as \newelsup, which is renormalizable. Alternatively,
one can proceed as follows. Consider first SQCD with vanishing
superpotential and $N_f<3N_c$. Due to asymptotic freedom, it approaches
a free fixed point in the UV, while at long distances it is governed by
a non-trivial fixed point, at which the scaling dimension of the meson
operator \defmes\ is smaller than its free field value. For $N_f<2N_c$,
the IR dimension is sufficiently small that the quartic perturbation in
\suppot\ becomes a relevant perturbation of this fixed point. Thus, in
this regime the theory with superpotential \suppot\ is well defined in
the UV. Our analysis of metastable states was performed (following
\IntriligatorDD) in the region $N_f<{3\over2}N_c$, in which the magnetic
theory is infrared free and the electric theory is UV complete.

A natural generalization of the problem studied here is to higher
order superpotentials of the form \elsuppot. The supersymmetric
ground states can be analyzed as in section 2. For example, in
the mesonic branch, in which the matrix $M$ is non-degenerate,
the F-term equation \nonsing\ takes in the general case the form
\eqn\fgen{\sum_{n=1}^{n_0} {1\over (n-1)!}m_n {\rm
Tr}M^n= \left(\det M\over\Lambda_e^{3N_c-N_f}\right)^{1\over
N_f-N_c}I_{N_f}~.}
Since the left hand side is a polynomial of
degree $n_0$, the matrix $M$ has at most $n_0$ distinct
eigenvalues, $(x_1,x_2,\cdots, x_{n_0})$. In a general vacuum, it
takes the form \eqn\formgen{M={\rm diag}(x_1^{l_1},
x_2^{l_2},\cdots, x_{n_0}^{l_{n_0}})} with $l_j=0,1,\cdots, N_f$
and $\sum_j l_j=N_f$. It seems clear that for large $N$ the number
of vacua grows like $N^{n_0}$, where $N$ denotes collectively
$N_f$, $N_c$ and any linear combinations of the two. The growth of
metastable states with $N$ is even faster since we can follow the
construction of section 3 in each block of $M$ \formgen\
separately.

{}From the perspective of effective field theory it is natural to
consider superpotentials with $n_0$ of order $N_f$, $N_c$. The
number of stable and metastable vacua grows in this case at least
as fast as $N!$. To ensure that these vacua are all
long-lived one needs to fine tune the couplings $m_n$ in \fgen\
accordingly. Conversely, if the couplings are generic, one expects
a distribution of lifetimes, so that at least some of the
non-supersymmetric vacua are long-lived.

An interesting\foot{E.g. for applications to phenomenology.}
generalization of the analysis of this paper is to a system in
which the $SU(N_f)_{\rm diag}$ symmetry of \elsuppot\ is gauged.
This leads to adjoint SQCD, which was studied using a
generalization of Seiberg duality in
\refs{\KutasovVE\KutasovNP-\KutasovSS}. One can use these results
as well as those of this paper to analyze the metastable states in
this system.

As mentioned in the introduction, SQCD with the superpotential
\suppot\ has a simple embedding in string theory, as a low energy
theory on a system of intersecting D-branes and $NS5$-branes. In
a companion paper \GiveonEW\ we describe this embedding and show that
much of the structure found in the gauge theory description in this
paper is nicely realized in the brane system.

\bigskip
\noindent{\bf Acknowledgements:}
We thank N. Seiberg for correspondence.
This work is supported in part by the BSF -- American-Israel
Bi-National Science Foundation. AG is supported in part by a
center of excellence supported by the Israel Science Foundation
(grant number 1468/06), EU grant MRTN-CT-2004-512194, DIP
grant H.52, and the Einstein Center at the Hebrew University. DK
is supported in part by DOE grant DE-FG02-90ER40560 and the
National Science Foundation under Grant 0529954. AG and DK
thank the EFI at the University of Chicago and Hebrew University,
respectively, for hospitality.

\listrefs
\end